\documentclass[aps,pra,twocolumn,showpacs]{revtex4}
\usepackage{amssymb}
\usepackage{bm}
\usepackage{graphicx}

\begin{document}

\title{Operator fidelity susceptibility: an indicator of quantum criticality}

\author{Xiaoguang Wang}
\affiliation{Department of Physics and Center of Theoretical and
Computational Physics, The University of Hong Kong, Pokfulam Road,
Hong Kong, China} \affiliation{Zhejiang Institute of Modern Physics,
Department of Physics, Zhejiang University, Hangzhou 310027, China}

\author{Zhe Sun}
\affiliation{Department of Physics and Center of Theoretical and
Computational Physics, The University of Hong Kong, Pokfulam Road,
Hong Kong, China} \affiliation{Zhejiang Institute of Modern Physics,
Department of Physics, Zhejiang University, Hangzhou 310027, China}

\author{Z. D. Wang}
\affiliation{Department of Physics and
Center of Theoretical and Computational Physics, The University of
Hong Kong, Pokfulam Road, Hong Kong, China}

\begin{abstract}
We introduce an operator fidelity and propose to use its
susceptibility for characterizing the sensitivity of quantum systems
to perturbations. Two typical models are addressed: one is the
transverse Ising model exhibiting a quantum phase transition, and
the other is the one dimensional Heisenberg spin chain with
next-nearest-neighbor interactions, which has the degeneracy. It is
revealed that the operator fidelity susceptibility is a good
indicator of quantum criticality regardless of the system
degeneracy.
\end{abstract}

\maketitle

\textit{Introduction}---There are two important concepts,
entanglement and fidelity
 in quantum information theory~\cite{Nielsen}. These two
concepts are closely related to each other. For instance,  fidelity,
which was first proposed as a tool for describing the stability of a
quantum system to perturbations~\cite{Peres}, may be used to
characterize quantum entanglement~\cite{fi-ent}. Notably,
fidelity has
 recently been used to characterize
quantum phase transitions (QPTs)~\cite{fqpt1,fqpt2,fqpt3,fqpt33}. On
the other hand, entanglement has also been employed to be an
indicator of QPTs in many correlated quantum
systems~\cite{eqpt1,eqpt2,eqpt3,eqpt4}.

How to characterize the stability of a quantum system to
perturbations is an important issue as there is no quantum
counterpart of the classical Lyapunov exponent. The Loshmidt
echo~\cite{echo} has been adopted as a measure of the system
stability against perturbations, which is introduced as follows. Let
operators $U_0$ and $U_1$ denote the time evolutions of Hamiltonians
$H_0$ and  $H_1$, where $H_1$ is slightly different from Hamiltonian
$H_0$ with $H_1-H_0=\epsilon V$ as a small perturbation. In this case, the operator $U_{\text{e}%
}=U_0^\dagger U_1$ is referred to as the echo operator, and the
absolute value of its expectation over a specific state
$|\psi\rangle$ is defined as the Loshmidt echo
\begin{equation}
L_{|\psi\rangle}=|\langle \psi|U_0^\dagger U_1|\psi\rangle|.
\end{equation}
This is just the fidelity amplitude. Obviously, it is
state-dependent, i.e., one has to  choose an initial state
(artificially in many cases) to evaluate its response to
perturbations. This scenario to characterize QPT has a serious
limitation, e.g., it can hardly be applied to a degenerated ground
state, which has been a great challenge for a long time.

In this Letter, mainly motivated by the above challenge, we
introduce a new kind of fidelity measure, called operator fidelity,
and propose for the first time to use its susceptibility for
characterizing the stability of quantum systems to perturbations. A
distinct and significant merit lies in that it is state-independent
and, in particular, is able to characterize the quantum criticality
regardless of degeneracy. To illustrate the feasibility and
reliability as well as the merit of the introduced operator fidelity
susceptibility, here we employ it to investigate two typical QPT
systems: the quantum Ising model and the Heisenberg model with
next-nearest-neighbor interactions. Indeed, this fidelity
susceptibility is able to serve as an indicator of QPT. In addition,
for comparison, we also consider the mixed state fidelity
susceptibility to address the quantum criticality with the
ground-state degeneracy.

We begin with the definition of operator fidelity. Let $\mathcal{H}
$ be a $d$-dimensional Hilbert space. All linear operators on
$\mathcal{H}$ are represented by $d\times d$ matrices and thus their
own may be considered to be vectors in an expanded $d^2$-dimensional
Hilbert space $\mathcal{H}_{\text{HS}}$. The inner product
$\mathcal{H}_{\text{HS}}$ is defined as the Hilbert-Schmidt product,
i.e., for operators $A$ and $B$, $\langle A|B\rangle=\text{Tr}(A^\dagger B)$%
. In this sense, any linear operators on $\mathcal{H}$ can be
considered as a state on $\mathcal{H}_{\text{HS}}$. Thus, the
fidelity of two states can naturally be generalized to the operator
level. For two unitary evolution operators $U_0$ and $U_1$ on
$\mathcal{H}$, the fidelity between them is defined as
\begin{equation}  \label{fff}
F^2=\frac{1}{d^2}|\text{Tr}(U_0^\dagger
U_1)|^2=|\overline{\text{Tr}}(U_0^\dagger U_1)|^2,
\end{equation}
where the averaged tracing operation is defined as
$\overline{\text{Tr}}\left( \right) =$Tr$\left( \right)/d $. It is
notable that  one may obtain the averaged Loshmidt
echo~\cite{ALecho1}-\cite{ALecho4} after averaging
$L_{|\psi\rangle}$ over all states on $\mathcal{H}$ with a Haar
measure, and the averaged Loschmidt echo and the operator fidelity
are essentially equivalent. Remarkably, the operator fidelity
involves not only the ground state, but also all eigenstates of the
system. It quantifies the difference between two unitary operators,
and is a conserved quantity under local operation in Hilbert space
$\mathcal{H}_{\text{HS}}$.

 We can rewrite the echo operator $U_{\text{e}}$ as~\cite{fqchaos1}
\begin{eqnarray}
U_{\text{e}} &=&1-i\epsilon \int_{0}^{t}V_{I}(t_{1})dt_{1}  \nonumber \\
&&-\epsilon
^{2}\int_{0}^{t}\int_{0}^{t_{1}}V_{I}(t_{1})V_{I}(t_{2})dt_{1}dt_{2}+O(%
\epsilon ^{3}),
\end{eqnarray}%
where $V_{I}(t)=\exp (iH_{0}t)V(t)\exp (-iH_{0}t)$ is the perturbation
operator in the interaction picture. After tracing, we have
\begin{equation}
\overline{\text{Tr}}(U_{\text{e}})=1-i\epsilon \overline{\text{Tr}}[W(t)]-\frac{\epsilon ^{2}}{2}%
\overline{\text{Tr}}[W(t)^{2}]+O(\epsilon ^{3}),
\end{equation}%
where $W(t)=\int_{0}^{t}V_{I}(t^{\prime })dt^{\prime }$. Then from
Eq.(2),  we obtain
\begin{equation}
F^{2}=1-\epsilon
^{2}[\overline{\text{Tr}}(W(t)^{2})-\overline{\text{Tr}}
^{2}(W(t))]+O(\epsilon ^{4}).  \label{ff}
\end{equation}%
 To evaluate the above operator fidelity, one has to choose a small
parameter artificially, which is $\epsilon $-dependent. To avoid
this artifact, we can also introduce a so-called fidelity
susceptibility~\cite{fqpt1,fqpt4}, which is given by
\begin{equation}
\chi _{_{F}}=\lim_{\epsilon \rightarrow 0}\frac{1-F}{\epsilon
^{2}}=\frac{1}{2}\{ \overline{\text{Tr}}\left[ W(t)^{2}\right]
-\overline{\text{Tr}}^{2}\left[ W(t)\right]\}.  \label{fs}
\end{equation}%
 Remarkably, the above simple formula possesses a
distinct computational advantage that enables one to calculate
straightforwardly the fidelity susceptibility from $W(t)$, which can
also be evaluated readily or at least numerically for more
complicated systems. On the other hand, generally speaking, a
quantity/measure susceptibility responds to the relevant
perturbations more sensitively than the quantity/measure itself
does, so we believe that it could capture a drastic change feature
of the quantum evolution (versus the relevant parameter) around a
critical point. We below explore the intriguing relationship between
the operator fidelity susceptibility and the QPT in two typical
systems, with one having degeneracy.

\textit{Quantum phase transition}---The first system we consider is
an Ising spin chain subject to a transverse magnetic field, whose
Hamiltonian reads
\begin{equation}
H_{0}=\sum_{l=-M}^{M}\left( \sigma _{l}^{x}\sigma _{l+1}^{x}+{\lambda }\frac{%
\sigma _{l}^{z}}{2}\right) ,
\end{equation}%
where ${\lambda }$ characterizes the strength of the transverse field, $%
\sigma _{l}^{\alpha }\left( \alpha =x,y,z\right) $ are the Pauli
operators defined on the $l$-th site, and the total number of spins
in the Ising chain is $N=2M+1$. The perturbation operator is given
by $\epsilon V=\epsilon \sum_{l=-M}^{M}{\sigma _{l}^{z}}/{2}$. There
are two competing terms in the Hamiltonian, i.e, the Ising
interaction and the transverse field term.

The Hamiltonian can be diagonalized by combining Jordan-Wigner
transformation and Fourier transformation to the momentum space,
i.e.,
\begin{equation}
H_0=\sum_{k>0}e^{i\frac{{\theta}_{k}}{2}\sigma _{kx}}\left( {\Omega}%
_{k}\sigma _{kz}\right) e^{-i\frac{{\theta}_{k}}{2}\sigma _{kx}}+\left( {1}-%
\frac{\lambda}{2}\right) \sigma _{0z}  \label{diag_H}
\end{equation}%
where we have used the following pseudospin operators $\sigma _{k\alpha
}\left( \alpha =x,y,z\right)$: $\sigma _{kx} =d_{k}^{\dagger
}d_{-k}^{\dagger }+d_{-k}d_{k},\left( k=1,2,...M\right), \sigma _{ky}
=-id_{k}^{\dagger }d_{-k}^{\dagger }+id_{-k}d_{k}, \sigma _{kz}
=d_{k}^{\dagger }d_{k}+d_{-k}^{\dagger }d_{-k}-1, \sigma _{0z}
=2d_{0}^{\dagger }d_{0}-1$. Operators $d_{k}^{\dagger },d_{k}\{k=1,2,...M\}$%
\ denote the fermionic creation and annihilation operators in the momentum
space. Here,

\begin{eqnarray}
{\Omega }_{k} &=&\sqrt{\left[ -\lambda {+}2\cos \left( \frac{2\pi k}{N}%
\right) \right] ^{2}+4\sin ^{2}\left( \frac{2\pi k}{N}\right) },  \nonumber
\\
{\theta }_{k} &=&\arcsin \left[ \frac{-2\sin \left( \frac{2\pi k}{N}\right)
}{{\Omega }_{k}}\right] .
\end{eqnarray}

Then the time evolution operator is derived as (with $\hbar =1$)
\begin{equation}  \label{uuu}
U_{0}(t)=e^{-i(-\frac{\lambda}{2}{+1)}\sigma _{0z}t}\prod_{k>0}e^{i\frac{{%
\theta}_{k}}{2}\sigma _{kx}}e^{-it{\Omega}_{k}\sigma _{kz}}e^{-i\frac{{\theta%
}_{k}}{2}\sigma _{kx}}.
\end{equation}
The unitary operator $U_1(t)$ for Hamiltonian $H_1=H_0+\epsilon V$
can be obtained by just replacing $\lambda$ with $\lambda+\epsilon$
in the above equation.

At this stage, from $U_0(t)$ and $U_1(t)$ given above, we are able
to obtain $W(t)$  as,
\begin{eqnarray}
W\left( t\right)  &=&\sum_{k>0}[\sigma _{kz}\left( t\cos ^{2}{\theta }_{k}+%
\frac{\sin ^{2}{\theta }_{k}}{2{\Omega }_{k}}\sin 2t{\Omega }_{k}\right)  \nonumber\\
&&+\sigma _{ky}\cos {\theta }_{k}\sin {\theta }_{k}\left( t-\frac{1}{2{%
\Omega }_{k}}\sin 2t{\Omega }_{k}\right)  \nonumber\\
&&+\sigma _{kx}\sin {\theta }_{k}\frac{1}{2{\Omega }_{k}}\left( \cos 2t{%
\Omega }_{k}-1\right) ]+\frac{{t}}{2}\sigma _{0z}
\end{eqnarray}
Consequently,
the fidelity susceptibility is derived exactly
\begin{eqnarray}
\chi _{_{F}} &=&\frac{1}{2}{t^{2}}\left( \frac{1}{2}+\sum_{k>0}\cos
^{2}\theta _{k}\right)   \nonumber \\
&&+\frac{1}{2}\sum_{k>0}{\sin ^{2}(\Omega _{k}t)\sin ^{2}\theta _{k}}/{%
\Omega _{k}^{2}.}  \label{Sus_F}
\end{eqnarray}%
Note that the first term in Eq.~(\ref{Sus_F}), which is proportional
to the square of time $t$,  plays a dominant role when $t$ is large.

In the transverse Ising model, two phases are separated by the
quantum phase transition point $\lambda=2$.
The singular behavior of QPT at the transition point reflects the
sensitivity of ground state to perturbations. At this stage, we
numerically look into the behaviors of the operator fidelity
susceptibility and its partial derivative with respect to $\lambda$
at a finite time $t=100$ (the natural units are used here). As shown
in Fig.~1 for different system sizes, the transition point is
unambiguously signatured: it is clearly seen that the fidelity
susceptibility and its partial derivative  are nearly unchanged when
increasing $\lambda $ from $0$ to $2$; the derivative increases
sharply at the transition point and the derivative peak is higher
when the system size becomes larger.

\begin{figure}[tbp]
\includegraphics[width=6 cm, clip]{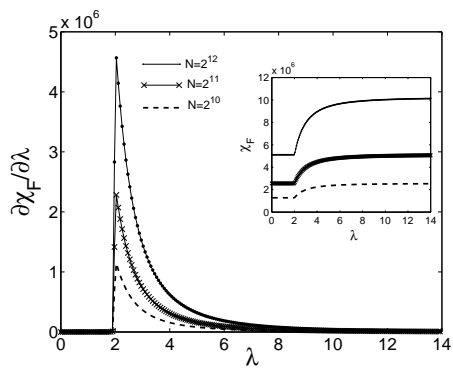}
\caption{Partial derivative of the fidelity susceptibility versus
the parameter $\protect\lambda$ for different system sizes $%
N=2^{10},2^{11},2^{12}$. The insert plots the fidelity
susceptibility $\protect\chi_F$ versus the parameter
$\protect\lambda$. The time $t=100$.}
\end{figure}

The behavior of the operator fidelity susceptibility at the QPT
point is different from that of the ground-state
fidelity~\cite{fqpt1}. The ground-state fidelity susceptibility
displays a sudden increase at the QPT point, reflecting the drastic
change of ground state of the system when the QPT occurs, while the
operator fidelity susceptibility drops to a certain value
continuously at the QPT point(and is unchanged below the point)
since it involves the all eigenstates and characterizes the
sensitivity of the whole system to perturbations in the time
evolution. However, on the other hand, its partial derivative
changes discontinuously at the QPT point and is much more sensitive
to perturbations, being able to single out the QPT point
unambiguously.

{\it Heiserberg model with next-nearest-neighbor interactions}---For
the fidelity scenario developed previously for QPTs, only pure
ground states can be addressed, without taking into account the
degeneracy; while it is the case for some quantum systems. As seen
above, the operator fidelity approach has an advantage that the
degeneracy is not necessary to be considered explicitly. To contrast
our approach with the state fidelity approach, we below address a
model with the ground-state degeneracy. The Hamiltonian of
one-dimensional Heisenberg system with next-nearest-neighbor
interaction reads
\begin{equation}
H_{0}=\sum\limits_{i}^{N}\left( J_{1}{\bf s}_{i}\cdot {\bf
s}_{i+1}+J_{2}{\bf s}_{i}\cdot {\bf s}_{i+2}\right) , \label{Hnnn}
\end{equation}%
where the ${\bf s}_{i}$ denotes the spin-1/2 operator at the
$i_{\text{th}}$ site, $N$ is the total number of sites,  $J_{1}$ and
$J_{2}$ are the nearest-neighbor (NN) and next-nearest-neighbor
(NNN) exchange couplings. As usual, we choose the periodic boundary
condition and  set $ J_{1}=1$ for convenience. The perturbation
operator is given by $\epsilon V=\epsilon \sum_{i}{\bf s}_{i}\cdot
{\bf s}_{i+2}$. Note that no exact analytical results are available
for this model (\ref{Hnnn}) except the special case of $J_{2}=0$ and
$J_{2}=1/2$.

It is well known that the point $J_{2}=1/2$ corresponds to the
Majumdar-Ghosh model where the ground state is the products of
dimers, leading to a gaped phase~\cite{NNNmodel}. Chen {\it et
al}~\cite{NNNfidelity} studied the ground-state fidelity and
first-excited-state fidelity of this system with even number of
sites. Here we focus on the  odd number of sites as the fourfold
degenerate energy level structure is present in this case.

 We first diagonalize the Hamiltonians numerically, and then calculate
 the operator fidelity susceptibility versus the NNN coupling $
J_{2}$ for $N=7,9,11$,  as plotted in Fig. 2(a). At a finite time
$t=100$, the susceptibility $\chi _{_{F}}$ decays to a minimum value
near the critical point $J_{2}=0.5$. With the size increasing, the
minimum point is closer to the critical point. It is expected that
the curve around the minimum point would become sharper and sharper
when the size increases, leading to a discontinuity in its partial
derivative with respect to $J_{2}$ at the critical point in the
thermodynamic limit, as in the case of the transverse Ising system.
We indeed note from the energy spectrum that the ground energy
level and the excited energy level crosses near the point $%
J_{2}=0.5$. In this sense, the operator-fidelity susceptibility (or
its partial derivative) is also able to capture the level crossing
feature in the system and thus
  to indicate the critical
point, overcoming the subtle problem induced by the degeneracy.

On the other hand, at least for comparison, it is also interesting
to consider an alternative approach to address degenerate cases by
making use of the mixed state fidelity given by~\cite{mixfidelity}
\begin{equation}
F\left( \rho _{0},\rho _{1}\right) \equiv \text{Tr}\left( \sqrt{\rho
_{1}^{1/2}\rho _{0}\rho _{1}^{1/2}}\right) =\text{Tr}\left( \sqrt{\rho
_{0}\rho _{1}}\right) .
\end{equation}%
Without loss of generality, it is not unreasonable to assume the
mixed ground state as an equal mixture of the degenerate ground
states,
\begin{equation}
\rho _{j}=\frac{1}{R}\sum\limits_{r=1}^{R}\left\vert \psi
_{jr}\right\rangle \left\langle \psi _{jr}\right\vert ,
\end{equation}%
with $r=1,2,...R$ denote the degeneracy and the state $\left\vert
\psi _{jr}\right\rangle $ denotes the $j_{\text{th}}$ degenerate
eigenstate of the system. In the fidelity $F\left( \rho _{0},\rho
_{1}\right) $ of this Heisenberg spin chain with the NNN
interactions, $\rho _{0}$ comes from\ the mixture of the ground
states of $H_{0}$, and $\rho _{1}$ corresponds to
$H_{1}=H_{0}+\epsilon V.$

\begin{figure}[tbp]
\includegraphics[width=4 cm, clip]{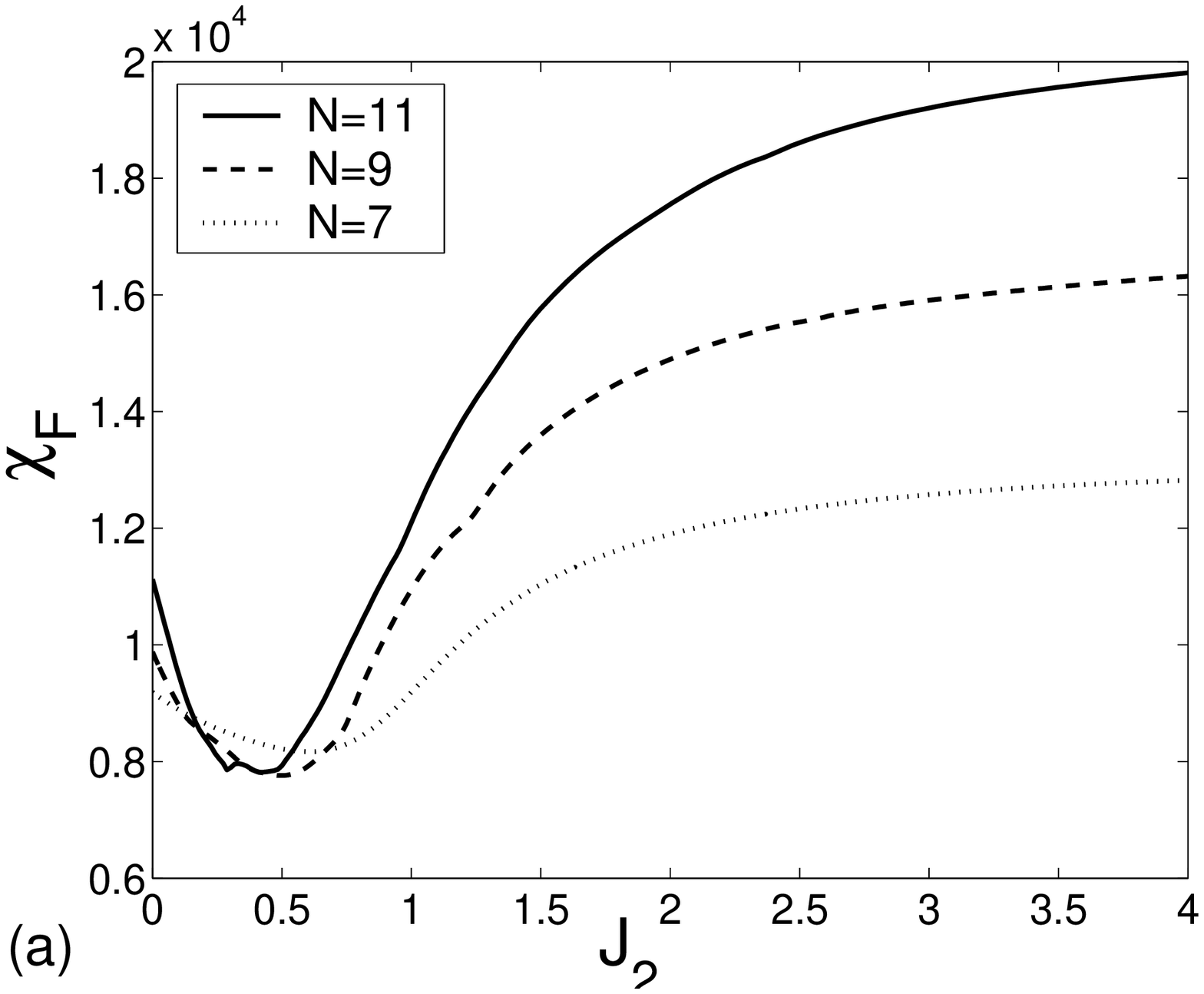}\hspace{0.5cm}
\includegraphics[width=4 cm, clip]{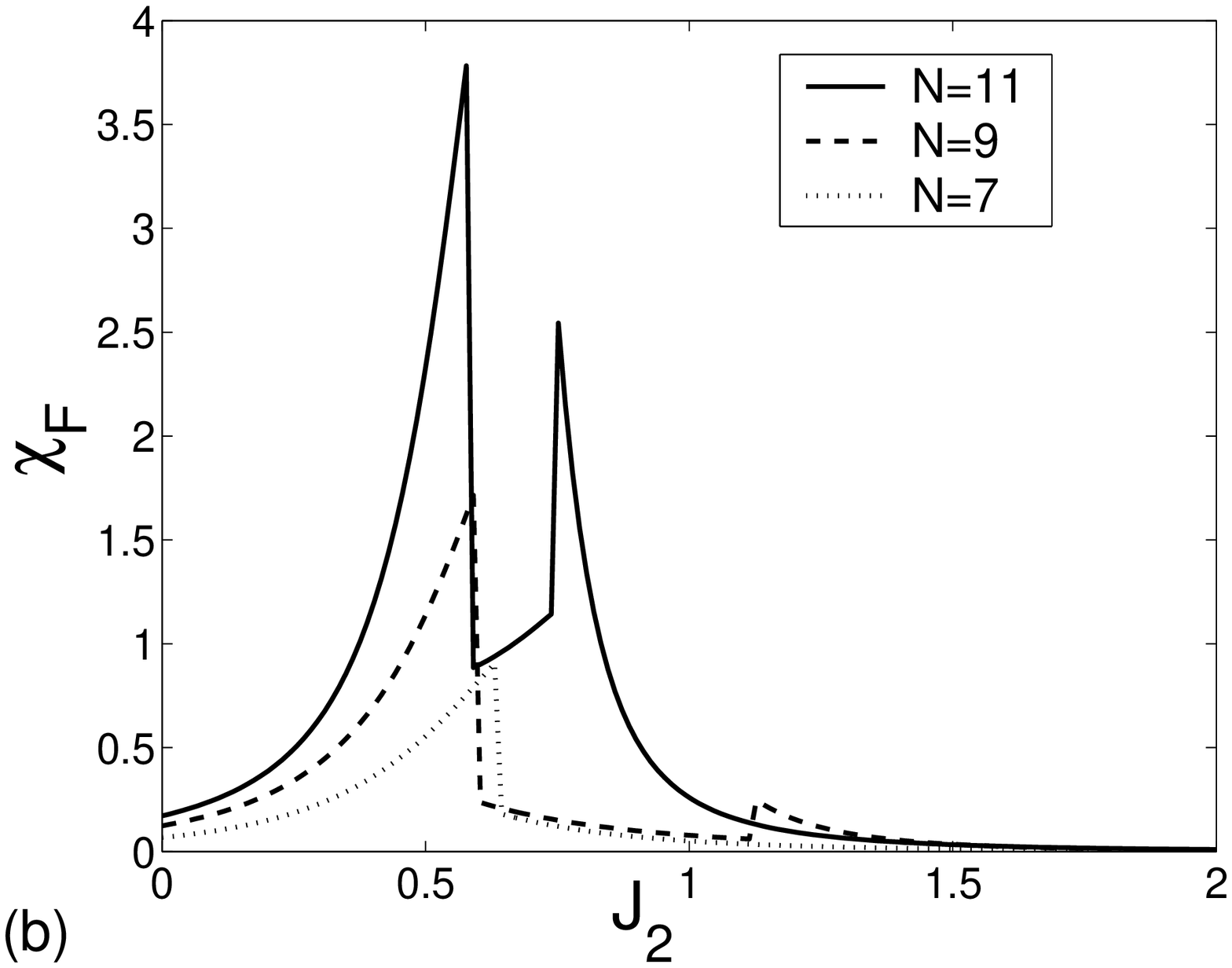}
\caption{(a) The operator fidelity susceptibility versus $J_2$ for
different system sizes $N=7,9,11$. (b) The mixed state fidelity
susceptibility versus $J_2$. }
\end{figure}

When the degeneracy of the system is explicitly obtained, we
evaluate the mixed state fidelity susceptibility versus the coupling
strength $J_{2}$ by combining Eq.(6) with Eq. (14), as shown in Fig.
2(b). Clearly, the susceptibility $\chi _{_{F}}$ passes the critical
point $J_{2}=0.5$ discontinuously. For larger sizes such as
$N=9,11$, there exist two peaks since the ground energy level
crossing occurs twice. With increasing the system size, the position
of the first peak approaches to the critical point, and the second
one is closer to the first one. Although, it seems that  the
suggested mixed state fidelity approach may also indicate the
critical point, it should be pointed out that it is feasible only
when the degeneracy of the ground states is explicitly known and the
equal mixture of the degenerated states is assumed. In addition, due
to the energy level crossing, the degeneracy may change at the
critical point and thus the mixed state fidelity approach may not be
workable for all the values of the considered parameter.

\textit{Relationship to entangling power}---Finally, we would like
to disclose an intrinsic connection between the present operator
fidelity and the entangling power~\cite{ep1} that was adopted to
characterize the entangling capability of a quantum evolution. The
entangling power is essentially the mean state linear entropy at
time $t$ after averaging over all initial states. We consider  a
general Hamiltonian in the form
\begin{equation}
H=I\otimes H_0+|1\rangle\langle 1|\otimes V=|0\rangle\langle 0|\otimes
U_0+|1\rangle\langle 1|\otimes H_1,
\end{equation}
where $I$ is the identity operator and $H_1=H_0+V$. The time
evolution operator is readily obtained as $U(t)=|0\rangle\langle
0|\otimes U_0(t)+|1\rangle\langle 1|\otimes U_1(t)$, where
$U_k(t)=\exp(-iH_kt)\{k=0,1\}$. This is a kind of the controlled-$U$
operator, for which the entangling power $e_{\text p}$ is
proportional to the operator entanglement $E[U(t)]$~\cite{ep2},
i.e., $e_{\text p}[U(t)]=[d/(d+1)]^2E[U(t)]$. The operator
entanglement $E[U(t)]$ can be straightforwardly obtained from the
expression of $U(t)$. Finally, we obtain
\begin{equation}
e_{\text
p}
=\frac{d^2}{2(d+1)^2}[1-F^2].
\end{equation} This establishes a direct connection between the
entangling power and operator fidelity.

\textit{Summary}--- We have proposed an operator fidelity approach
to characterize the stability of quantum system to perturbations,
which possesses a remarkable advantage that it is state-independent
and is able to characterize the quantum criticality regardless of
degeneracy. We have employed the approach
 to reveal successfully the QPT points in two typical systems: the
quantum Ising model and the Heisenberg chain with
next-nearest-neighbor interactions, with the latter having the
degeneracy. Our approach is quite promising for the exploration of
quantum instability including QPT and quantum chaos.

{\it Acknowledgements} The authors would like to thank C. P. Sun, S.
J. Gu, and P. Zanardi for their helpful discussions. The work was
supported by the Program for New Century Excellent Talents in
University (NCET), the NSFC with grant nos. 90503003 and 10429401,
the State Key Program for Basic Research of China with grant nos.
2006CB921206 and 2006CB921800, the Specialized Research Fund for the
Doctoral Program of Higher Education with grant No.20050335087, and
the RGC grants (HKU7045/05P and HKU7049/07P).

\end{document}